\documentclass[floats,twocolumn,prl,preprintnumbers,shortbibliography,10pt]{revtex4-1}
\usepackage{amsfonts,amsmath,graphicx,epsfig,amssymb}
\usepackage{epsfig}
\usepackage{mathrsfs}
\usepackage{latexsym}
\usepackage{amsxtra}
\usepackage{amsbsy}
\usepackage{amscd}
\usepackage{color}
\usepackage{bbm}
\usepackage{hyperref}
\usepackage{multirow}
\usepackage{graphics}
\usepackage{amsfonts}
\usepackage{subfigure}
\usepackage{color}
\allowdisplaybreaks

\newcommand{\be}{\begin{equation}}
\newcommand{\ee}{\end{equation}}
\newcommand{\sectitle}[1]{\vspace{.6cm}{\em #1.--}}

\def\no{\nonumber}
\def\bea{\arraycolsep .1em \begin{eqnarray}}
\def\eea{\end{eqnarray}}

\begin{document}
\title{Pole analysis on the doubly charmed meson in $D^0D^0\pi^+$  mass spectrum}
\author{Ling-Yun Dai$^{1,2}$}\email{dailingyun@hnu.edu.cn}
\author{Xiang Sun$^{1,2}$}
\author{Xian-Wei Kang$^{3,4}$}
\author{A. P. Szczepaniak$^{5,6,7}$}\email{aszczepa@indiana.edu}
\author{Jie-Sheng Yu$^{1,2}$}\email{yujiesheng@hnu.edu.cn}

\affiliation{$^{1}$\small School of Physics and Electronics, Hunan University, Changsha 410082, China\\
$^2$ Hunan Provincial Key Laboratory of High-Energy Scale Physics and Applications, Hunan University, Changsha 410082, China\\
$^3$ Key Laboratory of Beam Technology of the Ministry of Education, College of Nuclear Science and Technology, Beijing Normal University, Beijing 100875, China\\
$^4$ Beijing Radiation Center, Beijing 100875, China\\
$^5$ Physics Department, Indiana University, Bloomington, IN 47405, USA\\
$^6$ Center for Exploration of Energy and Matter, Indiana University, Bloomington, IN 47403, USA\\
$^7$ Thomas Jefferson National Accelerator Facility, Newport News, VA 23606, USA
}

\begin{abstract}
In this paper we study the scattering amplitudes of  $D^{0}D^{0}\pi^+$-$D^{*+}D^{0}$ coupled channels based on $K$-matrix
 within  the Chew-Mandelstam formalism. The $D^{0}D^{0}\pi^{+}$
invariant mass spectrum of LHCb is fitted and the pole parameters of the $T_{cc}^+$ are extracted. The analysis of pole behavior suggests  that the $T_{cc}^+$ may originate from a $D^{*+}D^{0}$ virtual state
and is formed as a result of  an interplay between an attractive interaction between $D^0$ and $D^{*+}$ and coupling to
 $D^{0}D^0\pi^+$  channel.
\end{abstract}
%

\maketitle

\parskip=2mm
\baselineskip=3.3mm

\sectitle{Introduction}
\label{sec:Intro}
For over half a century  the quark model
served as the fundamental template for constructing hadrons~\cite{Gell-Mann:1964ewy,Zweig:1964ruk}. Dozens of known  hadrons can be classified according to  this model with three quarks in a baryon and a quark-antiquark pair in a meson. However, the requirement of color neutrality alone does not preclude existence of  more complicated structures,
 including, for example tetraquarks and  pentaquarks.
In the last twenty years several candidates for such multi-quark hadrons, specifically containing heavy quarks,  have been observed by the Belle, BaBar, BESIII, D0, CDF, CMS, and LHCb
experiments\cite{Belle:2003nnu,CDF:2009jgo,Belle:2011aa,BESIII:2013ris,Liu:2013dau,BESIII:2013ouc,D0:2013jvp,CMS:2013jru,LHCb:2016axx,LHCb:2020bls,LHCb:2020pxc}.  Significant number of these states are found lying close to various thresholds for decays into non-exotic hadrons.
For example the $X(3872)$  discovered by Belle \cite{Belle:2003nnu}
 is in a mass region that is not expected to host a quark model-like charmonium state, but
 it is  only  $\sim 1\mbox{ MeV}$ away from the $D\bar{D}^{*}$ threshold. Proximity to this threshold makes it likely to be a $D\bar{D^{*}}$ molecule~\cite{Guo:2017jvc,Liu:2019zoy,Brambilla:2019esw}.
Recently, the LHCb Collaboration announced  observation of another $X$-like candidate, this time however, containing two charm quarks instead of a charm-anti-charm pair, labeled $T_{cc}^+$ \cite{LHCb:2021vvq,LHCb:2021auc}.
 The $T_{cc}^+$ was observed,  with a $21.7 \sigma$ significance, in the  $D^{0}D^{0}\pi^{+}$ invariant mass spectrum near threshold, {\it aka.} with the mass  close to the $X(3872)$, $M_{T_{cc}^+}-(M_{D^{*+}}+M_{D^{0}})=-237\pm61~keV/c^2$ and width, $\Gamma_{T_{cc}^+}=410\pm165~keV$. Because two charm quarks alone
cannot form a color singlet hadron, if confirmed, the $T_{cc}^+$ would be a clear evidence of a multi-quark hadron. The small width indicates that there could be a pole in the relevant partial wave close to the $D^0D^{*+}$ threshold,
However, since $D^*$  decays to $D\pi$, rescattering between $D^0D^0\pi^+$ and $D^0D^{*+}$ should be taken into account in determining the pole parameters~\cite{Dai:2012pb,Kang:2013jaa,Dai:2014zta,Danilkin:2014cra,Chen:2016mjn,Yao:2020bxx}.

There have been some theoretical studies for the $T_{cc}^+$, see e.g.\cite{Agaev:2021vur,Feijoo:2021ppq,Qin:2020zlg,Li:2021zbw,Dong:2021bvy,Ling:2021bir}. In this note we use effective range approximation  and consider the coupled amplitudes for production of  $D^{0}D^{0}\pi^+$ and $D^{*+}D^{0}$ final states.
By fitting to the line shape we obtain a solution for the production amplitude which enables analytical continuation to the complex energy plane where we extract the pole parameters. Finally, by analyzing the pole position we  speculate on the possible nature of the  $T_{cc}^+$ peak.

\sectitle{Formalism}\label{sec:formalism}
We need analytical amplitudes to describe the $D^0D^0\pi^+$ invariant mass spectrum in order to obtain accurate pole information.
The $T_{cc}^+$ is found in the $D^0D^0\pi^+$ invariant mass spectrum near the  $D^{*+}D^0$ threshold of  3875.09 MeV. One also notices that the branching ratio of $D^{*+}\to D^0\pi^+$ is $67.7\pm0.5\%$  \cite{ParticleDataGroup:2020ssz}.
The $D^*$ and $D\pi$ are physically two different  states. The former is a resonance (presumably a $q\bar q$ bound state in  quenched QCD) and the  later is in a two-hadron continuum. 
The two are distinguished,  for example by the value of the corresponding thresholds, which is relevant  given the proximity of the $T_{cc}$ to the $DD\pi$  threshold. 

Hence it is natural to consider the  $D^0D^0\pi^+$-$D^0D^{*+}$ coupled channels.
The analytical coupled channel amplitudes near threshold can be parametrized using a real, symmetric $2\times $2 $K$-matrix to describe the analytical part of the inverse amplitudes,
\bea
T^{-1}(s)= K^{-1}(s) -C(s) \,,\label{Eq:KM;T}
\eea
The matrix elements $C_i(s)$ of the diagonal $2\times 2$ Chew-Mandelstam (CM)  function \cite{Chew:1960iv,Edwards:1980sa,Kuang:2020bnk},   $C(s) = C_i(s) \delta_{i,j}$ contain the right hand cuts
  starting at the the thresholds, $s_{th,i} = (M_i + m_i)^2$.  Here the masses are $M_1=M_{D^0}+m_{\pi^+}$, $m_1=M_{D^0}$, and  $M_2=M_{D^{*+}}$,  $m_2=M_{D^0}$ for the
  $D^0D^0\pi^+$ and $D^0D^{*+}$ channels, respectively.
  Note that the $D^0\pi^+$ system is treated as an isobar of spin-1 and therefore the $T$ describes $S$-wave amplitudes. The CM function,
 \bea
C_i(s)=\frac{s}{\pi}\int_{s_{th_i}}^{\infty} ds'\frac{\rho_i(s')}{s'(s'-s)},\label{eq:C}
\eea
is defined by the  (quasi)two-body $S$-wave phase space factor, $Im C_i(s) = \rho_i(s) = \lambda^{1/2}(s,M_i^2,m_i^2)/s$, explicitly,
\bea
&&C_i(s;M_i,m_i)=\left[\frac{M_i^2-m_i^2}{\pi s}-\frac{M_i^2+m_i^2}{\pi (M_i^2-m_i^2)} \right] \ln\left(\frac{m_i}{M_i}\right)   \no\\
&&+\frac{1}{\pi}+\frac{\rho_i(s)}{\pi}\ln\left[\frac{\sqrt{s_{th_i}-s}-\sqrt{(M_i-m_i)^2-s}}{\sqrt{s_{th_i}-s}+\sqrt{(M_i-m_i)^2-s}}\right]. \label{Eq:C;ana}
\eea
With the threshold singularities accounted  for by $C(s)$, the $K$-matrix is analytical in the vicinity at thresholds and in the effective range approximation it as approximated by a matrix of constants.

In the notation of  \cite{Au:1986vs,Dai:2014lza} the $s$-dependence of production amplitude for the processes $pp\to D^0D^0\pi^+ +X$ and $pp\to D^0D^{*+} +X$,
can be represented  by a two-dimensional  vector
\bea
F_i(s)&=&\sum_{k=1}^2\alpha_k(s) T_{ki}(s) \; . \label{eq:F}
\eea
where $\alpha_i(s)$ are regular functions of $s$ on the physical cut. Since the range of invariant mass is small, with $\Delta \sqrt{s} << O(100 \mbox{ MeV})$, we can safely ignore any variation in $s$ of the production amplitudes, $\alpha_i(s)$  and also approximate them by  constants.
Finally, the measured yield is proportional to the differential cross section and given by
\bea
\frac{dY_1}{d\sqrt{s}}=\; N p_1 |F_1|^2. \label{eq:dGds}
\eea
Here $p_1= \lambda(s,M_1^2,m^2_1)/2\sqrt{s}$ is the magnitude of the momentum of the $\pi^+$ in the center of mass frame.
Since the overall number of events  is fitted we can absorb  $\alpha_1$ into the normalization factor $N$ and thus we set $\alpha_1=1$.

\begin{figure}[hpt]
\includegraphics[width=0.48\textwidth,height=0.25\textheight]{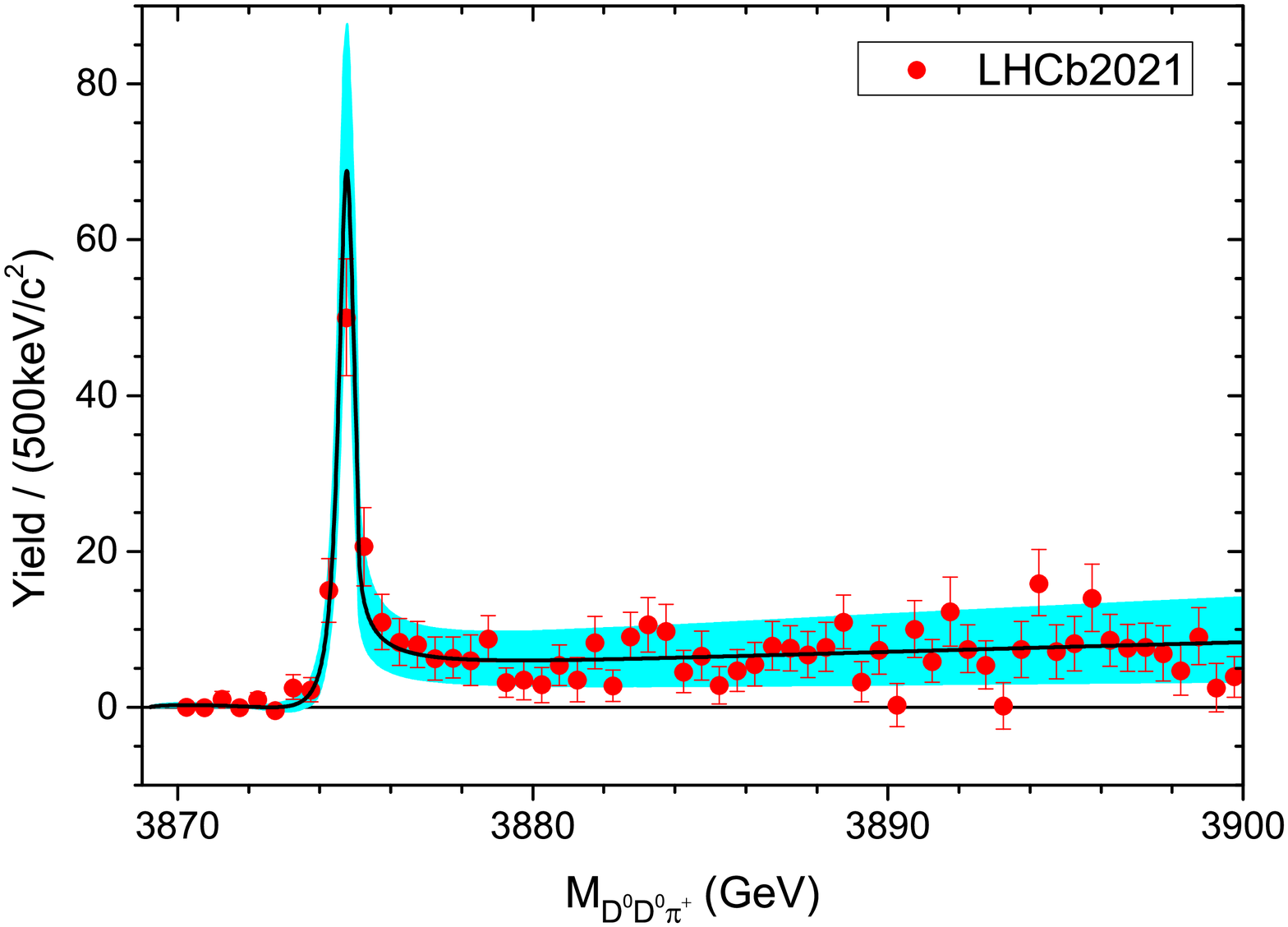}
\includegraphics[width=0.48\textwidth,height=0.25\textheight]{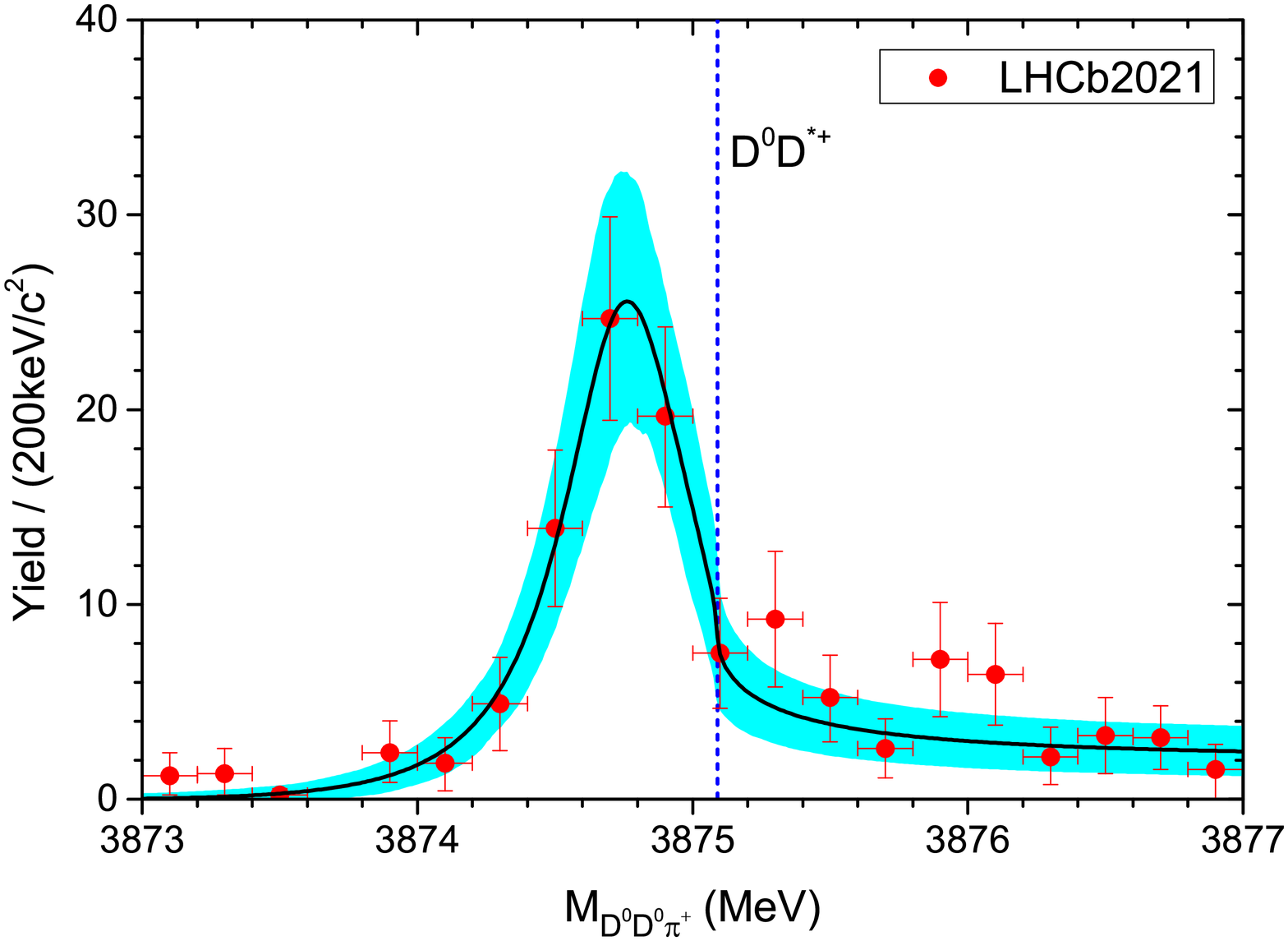}
\caption{\label{Fig:events;2}The clear line shape from our solutions.  The data points of $D^0D^0\pi^+$ invariant mass spectrum are superimposed.
The cyan bands are taken from bootstrap method within 2$\sigma$. } 
\end{figure}

\sectitle{Fit results and discussion}\label{sec:fit}
We fit the amplitudes to the $D^0D^0\pi^+$  invariant mass spectrum \cite{LHCb:2021vvq,LHCb:2021auc}  using MINUIT \cite{James:1975dr}.
One needs to consider the resolution for the $D^0D^0\pi^+$ mass. Here we follow the experiment \cite{LHCb:2021auc}, convolute Eq.~(\ref{eq:resolution}) with the resolution function. For a data point with mass $E_i$, we get the Yields for the bin:
\bea
\frac{{\rm Yields}}{{\rm \Delta E}}&=& \int_{(E_i-\Delta E/2)^2}^{(E_i+\Delta E/2)^2} d s~  \frac{N p_{1} |F_1|^2}{2\Delta E\sqrt{s}}
  \no\\
&&\left\{\sum_{j=1}^{2}\beta_j\exp\left[-\frac{1}{2}\left(\frac{\sqrt{s}-E_i}{\sigma_j}\right)^2\right]\right\}\,, \label{eq:resolution}
\eea
where $\Delta E$ is the bin width, $\beta_1=0.778$, $\beta_2=0.222$, $\sigma_1=1.05\times263$~keV, and $\sigma_2=2.413\times\sigma_1$ \cite{LHCb:2021auc}.  
We find a unique solution with desired physical properties.
 The parameters of the fit are given in Table \ref{tab:para}, and correspond  to $\chi^2_{\rm d.o.f}=0.92$. Notice that the error of the parameters from MINUIT is much smaller than that from bootstrap \cite{Efron:1979bxm}, which is done by varying the data with experimental uncertainty multiplying a normal distribution function.
\begin{table}[h!]
\begin{center}
\begin{tabular}{l}
\hline\hline
$K_{11}=-0.01204\pm0.00691^{+0.03280}_{-0.07039}$ \\
$K_{12} = K_{21} =0.5080\pm0.0025^{+0.0348}_{-0.0700}$ \\
$K_{22} =1.4447\pm0.0015^{+0.0235}_{-0.0477}$ \\
$\alpha_2=-0.3024\pm0.0016^{+0.0261}_{-0.0583}$    \\
$N_a= 1434.0\pm 129.8^{+662.0}_{-964.8} \mbox{ GeV}^{-2}$   \\
$N_b=516.0\pm49.3^{+225.6}_{-363.4} \mbox{ GeV}^{-2}$    \\
$\chi ^2_{{\rm d.o.f}}$=0.92   \\[1mm]
\hline
\end{tabular}
\caption{\label{tab:para} Parameters of the best fit, as explained in the text. The $K$-matrix elements and production parameters,  $\alpha_i$ are  dimensionless.
The first uncertainty of the parameters is given from MINUIT, and the second (up and down) uncertainty is from bootstrap within 2$\sigma$. $N_a$ is the normalization factor for the data in the full range, while $N_b$ is for the data in the region of $T_{cc}^+$.  The correlation matrix is given in supplemental material.   }
\end{center}
\end{table}
The comparison between the data and the model is shown in Fig.~\ref{Fig:events;2}.
As can be seen, our amplitudes fit the data rather well.
To study the resonance we also enlarge the plot of our solution around the $T_{cc}^+$, multiplying $N_b$ instead of $N_a$ in Eq.~(\ref{eq:dGds}), as shown in the bottom graph in Fig.~\ref{Fig:events;2}.
Once the $D^0D^0\pi^+$-$D^0D^{*+}$ amplitudes are determined on the real axis  they can be  analytically continued to  extract the information about the singularities
 located on the nearby Riemann sheets (RS). These are reached
 from the real $s$-axis though the unitary cuts of the $C_i(s)$ functions.
Near the pole $s_R$ residues/couplings in $n$-th RS are computed from
\be
T^n_{ij}(s)\simeq \frac{g^n_i g^n_j}{s_R^n-s}\,.\label{eq;g}
\ee
We find a single pole on RS-II and the pole parameters are given in Table~\ref{tab:poles}.
\begin{table}[t]
\begin{center}
{\footnotesize
\begin{tabular}{|c|c|c|c|c|}
\hline
\rule[-0.2cm]{0cm}{7mm}\multirow{2}{*}{ pole location (MeV)}
& \multicolumn{2}{c|}{$g^{II}_{D^0D^0\pi^+}=|g|e^{i\varphi}$}  & \multicolumn{2}{c|}{$g^{II}_{D^0D^{*+}}=|g|e^{i\varphi}$}\\
\cline{2-5}
\rule[-0.2cm]{0cm}{7mm} & $|g_1|$~(GeV) & $\varphi_1$~($^\circ$) & $|g_2|$~(GeV) & $\varphi_2$~($^\circ$)   \\
\hline
\rule[-0.2cm]{0cm}{7mm} $3874.74^{+0.11}_{-0.04} -i~0.30^{+0.05}_{-0.09}$   &  $0.22^{+0.03}_{-0.04}$ & $9^{+11}_{-5}$     &   $0.69^{+0.04}_{-0.02}$ & $10^{+11}_{-5}$    \\
\hline
\end{tabular}
\caption{\label{tab:poles} The pole location and its residues (both magnitude and phase) from our fit, in RS-II. } 
}
\end{center}
\end{table}
Here we follow the standard labeling of the sheets, {\it e.g.} the second sheet is reached from the physical region by moving into the lower complex plane between two thresholds \cite{Frazer:1964zz}.
The uncertainty of the pole parameters come from bootstrap within 2$\sigma$.

Notice that in the bootstrap method, where the data points are varied randomly, all the poles are located in RS-II.
Since $|g_2| > |g_1|$ it appears that $T_{cc}^+$ couples more strongly (roughly a factor of three) to the $D^0D^{*+}$ channel than to the $D^0D^0\pi^+$ channel. This  supports the hypothesis that  $T_{cc}^+$ is a composite object dominated by the $D^0D^{*+}$ component.

\begin{figure}[ht]
\includegraphics[width=0.48\textwidth,height=0.25\textheight]{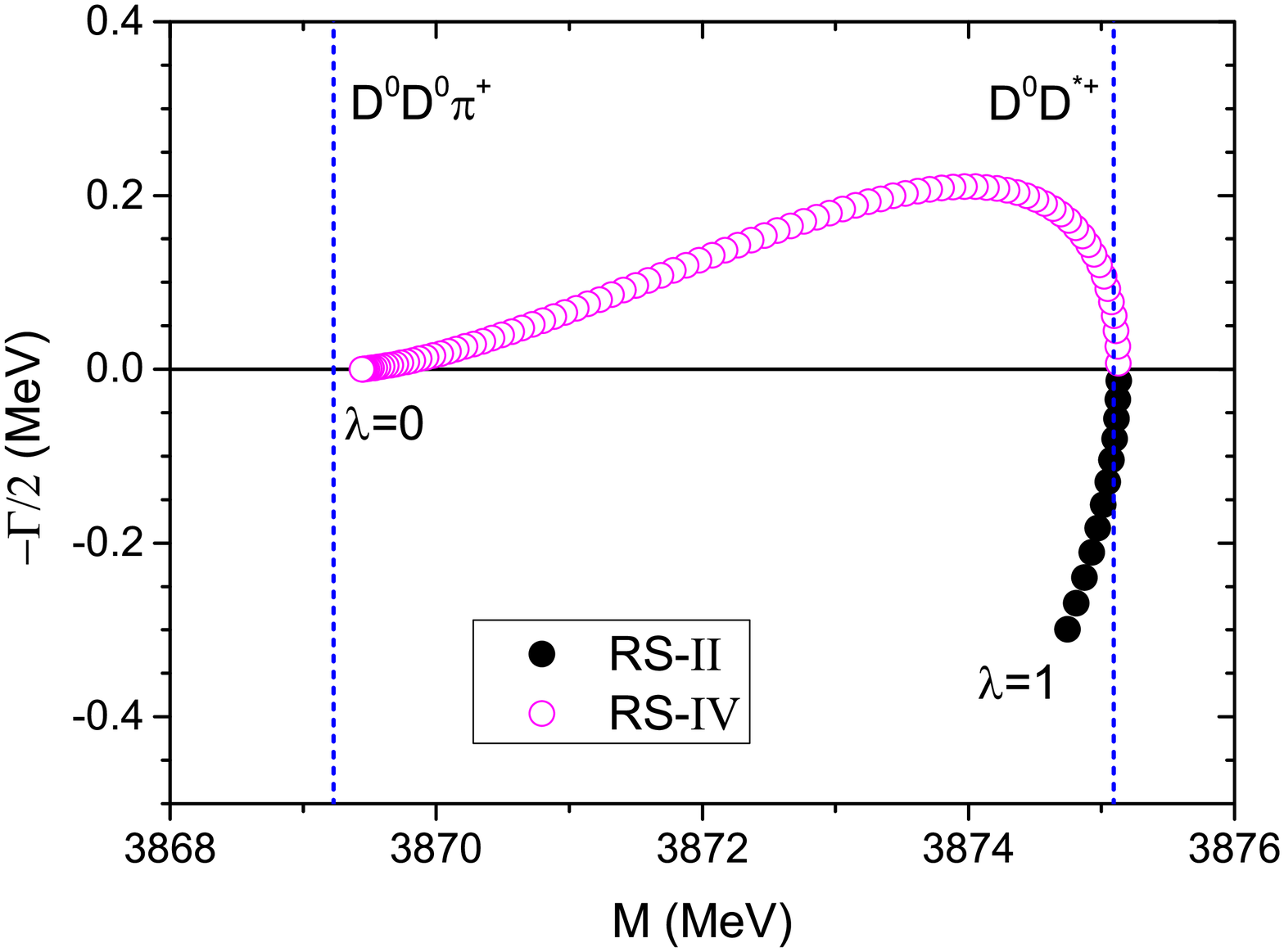}
\caption{\label{Fig:poletraj} The trajectories of pole locations by varying $\lambda$. The black filled circles are the poles in the second Riemann sheet,  $1 \geq  \lambda \geq  0.88 $,  and the magenta open circles are the poles in the fourth Riemann sheet, $ 0.87 \geq \lambda \geq 0  $, respectively.
The step of $\Delta \lambda$ is 0.01. }
\end{figure}

The trajectory of the pole on the second Riemann sheet is studied  by varying $\lambda$ which is introduced to modify the strength of coupling between the two channels, $K_{12}(s) \to\lambda K_{12}(s)$, so that $\lambda=1$ corresponds to the physical amplitude  while for $\lambda=0$, Eq.(\ref{Eq:KM;T}) represents two  uncoupled channels.
As a function of  $\lambda$ the pole trajectory is shown in Fig.~\ref{Fig:poletraj}.
As $\lambda$ decreases, the pole moves upwards from the lower  half $\sqrt{s}$-plane of RS-II to the upper half plane of RS-IV, crossing  the real axis above the  second (heavier) $D^0 D^{*+}$ threshold. This does not violate unitarity since while moving from the second to the  fourth sheet the pole never crosses the physical region. As this happens the resonance bump seen on the real axis between thresholds moves towards the heavier threshold  and  as the pole enters the fourth sheet it becomes a cusp.
 As $\lambda$ is decreased further the poles moves below the lower threshold and into the real axis.  Finally it reaches the mass just $\sim 0.21 \mbox{ MeV}$  above the $D^0D^0\pi^+$ threshold.
 Notice that at the end of the trajectory with $\lambda=0$, corresponding to the $D^0D^{*+}$ single channel, the pole in the real axis below threshold  is a virtual state.
 This implies that in the absence of channel coupling the $D^0D^{*+}$ system may not be sufficiently attractive to produce a molecule. If so  the $T_{cc}^+$ is not a true bound state (a pole remaining on the second sheet) but an effect of a complicated interplay of weak attraction and channel interactions.
This behaviour is  similar to that of the $P_c(4312)$, which was found to be likely an effect of weak interaction between $\Sigma_C^+ \bar D^0$ and coupling the the $J/\psi p$ channel \cite{Fernandez-Ramirez:2019koa}.

To further assess systematic uncertainties we include higher order terms in the effective range expansion. This slightly improves the fit quality but does not qualitatively change the amplitudes in the vicinity of the $T_{cc}^+$.
In yet another check, we include the fit to the data without the resolution function. In this case, still only one pole is found in the RS-II, $3874.75^{+0.12}_{-0.06} -i~0.34^{+0.06}_{-0.12}$~MeV, with the residues of $|g_1|=0.23^{+0.03}_{-0.04}$~GeV and $|g_2|=0.70^{+0.03}_{-0.03}$~GeV. Quite the same as what is found in Table~\ref{tab:poles}.

\sectitle{Role of the  $D^*$ width}
Since $D^{*+}$ is unstable its contribution to the spectral function corresponds to a branch cut (below the real  axis) and  not a pole. To account for that we modify the Chew-Mandelstam $C_2$ accordingly  \cite{Basdevant:1978tx}:
\bea
C_2(s) \to \frac{1}{\pi}\int_{s_{tr,D\pi}}^{\infty}ds' C(s;\sqrt{s'},m_2){\rm Im} f_{D\pi}(s')
 \eea
with $s_{th,D\pi} = (M_{D^0} + m_\pi^+)^2$ being the threshold for the reaction $D^0 \pi^+ \to  D^{*+} \to D^0\pi^+$ and  $f_{D\pi}(s) = D^{-1}(s)$,
the scattering amplitude in the single resonance approximation,  with
$D(s) = \tilde{M}^2 - s - \Sigma(s)$, and  $\Sigma(s)=g^2 (s-s_{th,D\pi})C(s;M_2,m_2)$ so that the  imaginary part,  ${\rm Im} \Sigma(s) = g^2 (s-s_{th,D\pi}) \rho_2(s)$ is the energy dependent width corresponding to the
$P$-wave decay of the $D^{*+} \to D^0 \pi^+$.
 With the parameters $g=0.4451$ and $\tilde{M}=2010.77$~MeV the  amplitude $f$ reproduces the line shape of the $D^*$
  correspond to a Breit-Wigner resonance with pole at
    $M_{D^{*+}}-i \Gamma_{M_{D^{*+}}}/2=2010.26-i0.04$~MeV.
With this modification of $C_2(s)$, we fit the coupled channel amplitudes again, convoluting with Eq.~(\ref{eq:resolution}) with the resolution function, and
find a rather similar solution to the previous one. A pole is found in RS-II with $3874.76^{+0.08}_{-0.04} -i~0.26^{+0.02}_{-0.09}$~MeV, and the residues are extracted as $|g_1|=0.21^{+0.01}_{-0.04}$~GeV and $|g_2|=0.71^{+0.02}_{-0.02}$~GeV. The pole trajectory is the same as what we found in Fig.\ref{Fig:poletraj}, with only the pole moves towards but not reach the real axis. Also the destination of the pole (with $\lambda=0$) being roughly 2.6~MeV above the $D^0D^0\pi^+$ threshold.
These support the conclusion before.

\sectitle{Summary}
In this letter we performed  amplitude analysis on the invariant mass spectrum of $D^0D^0\pi^+$. The $D^0D^0\pi^+$ ~-~ $D^0D^{*+}$ coupled channel scattering amplitude is constructed using a K-matrix within the  Chew-Mandelstam formalism.
Then we apply the Au-Morgan-Pennington method to study the final state interactions for the invariant mass spectrum of $D^0D^0\pi^+$.
A high quality fit to the experiment data of LHCb~\cite{LHCb:2021vvq,LHCb:2021auc} is obtained.
We find a pole in the second Riemann sheet for the $T_{cc}^+$, with the pole location $3874.74^{+0.11}_{-0.04} -i~0.30^{+0.05}_{-0.09}$~MeV.
By reducing the strength of inelastic channels we obtain the pole trajectory that suggests  $T_{cc}^+$ might be  a $D^0D^{*+}$ virtual state. Precise measurements of the line shape would be need to further reduce theoretical uncertainties.

\sectitle{Acknowledgements}
We thank C. Fern\'andez-Ram\'\i{}rez for helpful discussions on bootstrap method. 
This work is supported by Joint Large Scale Scientific Facility Funds of the National Natural Science Foundation of China (NSFC) and Chinese Academy of Sciences (CAS) under Contract No.U1932110, National Natural Science Foundation of China with Grants No.11805059, No.11805012, No.11675051 and No.12061141006, Fundamental Research Funds for the central universities of China, and U.S. Department of Energy Grants No. DE-AC05-06OR23177 and No. DE-FG02- 87ER40365.

\bibliography{ref}

\clearpage
\appendix
\setcounter{equation}{0}
\setcounter{table}{0}
\setcounter{figure}{0}
\renewcommand{\theequation}{\Alph{equation}}
\renewcommand{\thetable}{\Alph{table}}
\renewcommand{\thefigure}{\Alph{figure}}

\onecolumngrid
\section{Supplemental material}
\sectitle{Different fits}
To study systematic uncertainties we vary the number of parameters in the $K$-matrix,
\bea
K^{ij}(s)&=&\sum_{l}\frac{f^{i}_{l}f^{j}_{l}}{(s_{l}-s)}
+\sum_{n=0}c^{ij}_{n}(\frac{s}{s_{th2}}-1)^{n}\;.\label{eq;K}
\eea
Near  threshold the $K$-matrix can be simplified to effective range formula so  
  that  $f^i_l$, and all higher order polynomials, $c^{ij}_{n\geq 2}$ given in Eq.(\ref{eq;K}) are set to zero.  
In an alternative fit, which we refer to as  Fit A, we add one more term, proportional to $c^{ij}_{1}$,  in each element of the $K$. Thus Fit A, has
 six parameters to parametrize  the $T$-matrix, compared to three used in the nominal fit.
To reduce the correlation between parameters, in Fit A we use the same normalization  factors as obtained from the nominal fit presented in the paper.
Furthermore, to investigate the  effects of resolution, in Fit B we use the nominal model removing the smearing.
The parameters and $\chi^2/d.o.f$ of these fits   are summarized in Table~\ref{tab:para;check}.
\begin{table*}[h!]
{\footnotesize
\begin{center}
\begin{tabular}{|c||c|c|}
\hline
                & Fit A                     & Fit B                         \\
\hline\hline
$c^{11}_{0}$    & $-0.1067\pm0.0170^{+0.1079}_{-0.0415}$       & $0.007905\pm0.006404^{+0.034727}_{-0.082606}$          \\
$c^{11}_{1}$    & $10.5612\pm5.1641^{+10.4981}_{-4.0526}$                                 & $\cdots$         \\
$c^{12}_{0}$    & $0.5080$       & $0.5315\pm0.0025^{+0.0320}_{-0.0666}$        \\
$c^{12}_{1}$    & $-15.3292\pm3.2335^{+17.2790}_{-5.1199}$                                  & $\cdots$       \\
$c^{22}_{0}$    & $1.4550\pm0.0021^{+0.0145}_{-0.0054}$      & $1.4271\pm0.0016^{+0.0222}_{-0.0491}$        \\
$c^{22}_{1}$    & $8.3280\pm5.6488^{+49.6034}_{-17.3702}$                                 & $\cdots$    \\
$\alpha_2$      & $-0.2866\pm0.0027^{+0.0174}_{-0.0062}$         & $-0.3215\pm0.0015^{+0.0270}_{-0.0606}$     \\
$N_{a}$ ($\mbox{ GeV}^{-2}$)         & $1434.0$       & $1339.6\pm 117.7^{+639.1}_{-952.5}$     \\
$N_{b}$ ($\mbox{ GeV}^{-2}$)        & $516.0$        & $568.5\pm52.0^{+268.4}_{-394.8}$     \\
$\chi ^2_{{\rm d.o.f}}$  & 0.86                   & 0.95                  \\[1mm]
\hline\end{tabular}
\caption{\label{tab:para;check} Results of Fits A and B, as explained in the text. The $c^{ij}_0$ is dimensionless and the $c^{ij}_1$ is in unit of GeV$^{-2}$. The uncertainty of the parameters is given from MINUIT, and the upper and lower errors are taken from Bootstrap method.  }
\end{center}
}
\end{table*}
The quality of all fits are similar.
For completeness, in Table \ref{tab:cor} we give the correlation matrix between the parameters of the nominal fit.
\begin{table*}[h!]
{\footnotesize
\begin{center}
\begin{tabular}{|c|c c c c c |}
\hline
   &  $K_{11}$  & $K_{12}$  & $K_{22}$  & $\alpha_{2}$&  $N_a$      \\
\hline
$K_{11}$        & 1.000  & 0.422 & -0.067  & -0.580 &-0.082        \\
$K_{12}$        & 0.422  & 1.000 &-0.100   &-0.628  & -0.095     \\
$K_{22}$        &-0.067  &-0.100 &1.000    & 0.671  & 0.018        \\
$\alpha_{2}$    & -0.580 &-0.628 &0.671    & 1.000  &-0.012        \\
$N_a$           & -0.082 &-0.095 & 0.018   & 0.012  &1.000         \\
\hline\end{tabular}
\caption{\label{tab:cor} Correlation coefficients between parameters of the nominal fit using bootstrap.  $N_b$ is rescaling from $N_a$ and thus we do not list the correlation coefficients related to it.   }
\end{center}
}
\end{table*}

\begin{figure}[hpt]
\includegraphics[width=0.48\textwidth,height=0.25\textheight]{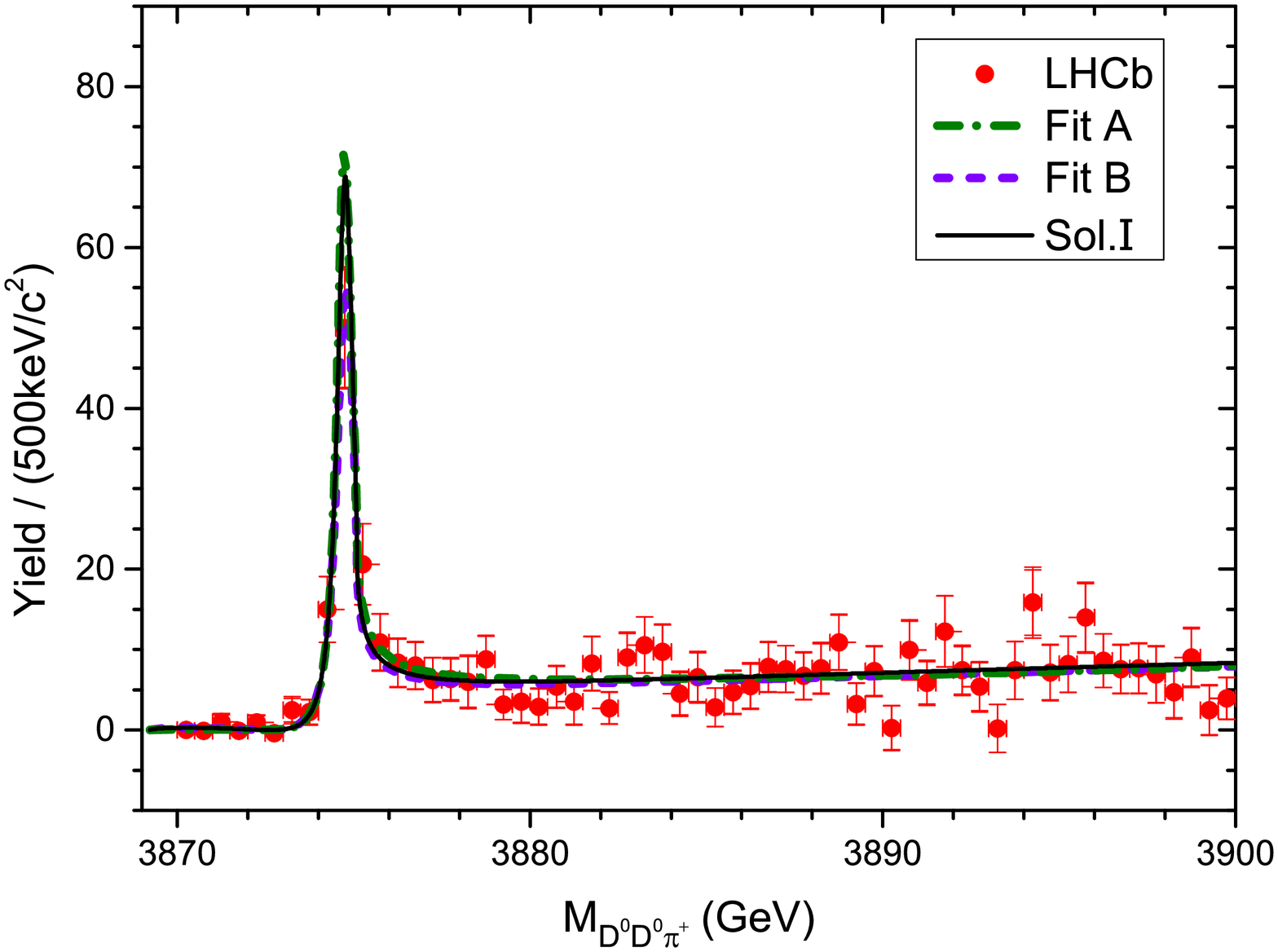}
\includegraphics[width=0.48\textwidth,height=0.25\textheight]{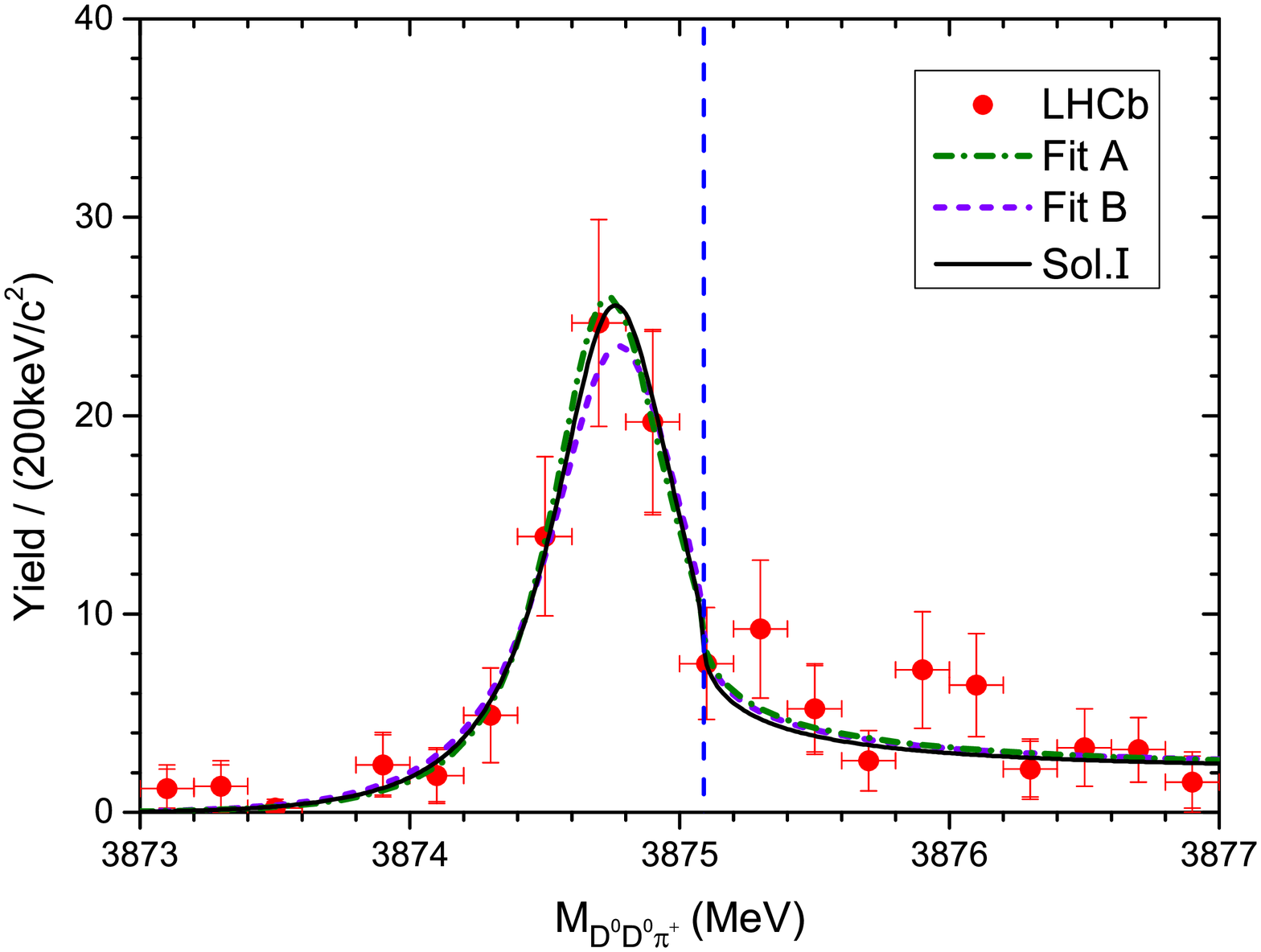}
\caption{\label{fig:events;check} Line shapes from our solutions.  
The olive dash-dotted lines are for Fit A and the violet dashed lines are for Fit B. }
\end{figure}

The larges deviation with respect to the nominal (referred to in the figures as Sol.~I)  is observed  in Fit B.
In Fit B, the peak appears to be a bit lower compared with the unresolved data. 
The mass distributions from the fits are shown in Fig.~\ref{fig:events;check} and in  Table~\ref{tab:poles;all} we give the $T_{cc}^+$ pole parameters.

\begin{figure}[ht]
\includegraphics[width=0.32\textwidth,height=0.18\textheight]{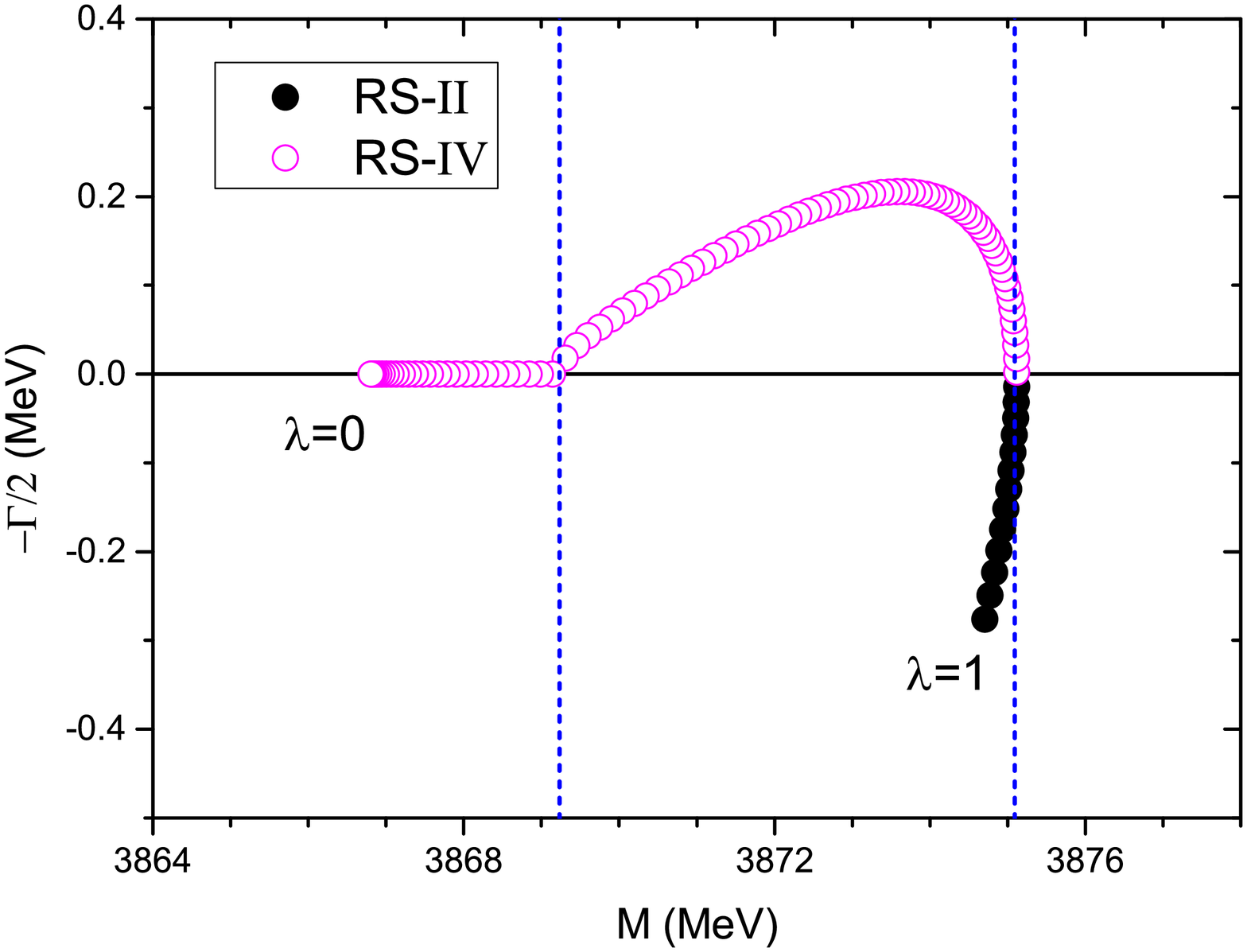}
\includegraphics[width=0.32\textwidth,height=0.18\textheight]{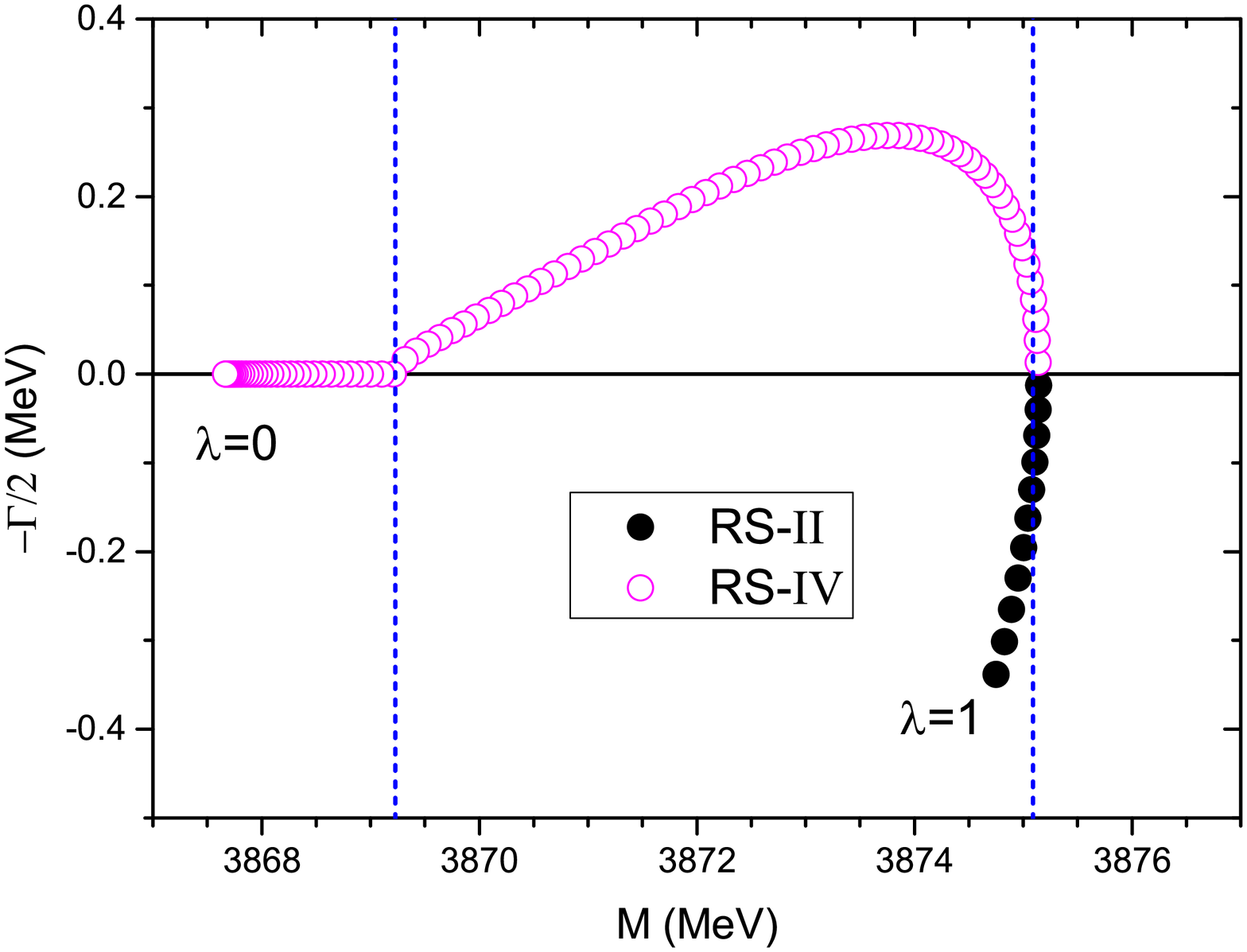}
\caption{\label{Fig:poletraj;all} The trajectories of pole locations by varying $\lambda$. The black filled circles  represent for the poles in the second Riemann sheet, and the magenta open circles represent for the poles in the fourth Riemann sheet. The step of $\Delta \lambda$ is 0.01. The left graph corresponds to Fit A, and the right one to Fit B. }
\end{figure}

\begin{table}[h!]
\begin{center}
{\footnotesize
\begin{tabular}{|c|c|c|c|c|c|}
\hline
\rule[-0.2cm]{0cm}{7mm}\multirow{2}{*}{ Cases}
&\multirow{2}{*}{ pole location (MeV)}
& \multicolumn{2}{c|}{$g^{II}_{D^0D^0\pi^+}=|g|e^{i\varphi}$}  & \multicolumn{2}{c|}{$g^{II}_{D^0D^{*+}}=|g|e^{i\varphi}$}\\
\cline{3-6}
\rule[-0.2cm]{0cm}{7mm} &  & $|g_1|$~(GeV) & $\varphi_1$~($^\circ$) & $|g_2|$~(GeV) & $\varphi_2$~($^\circ$)   \\
\hline
\rule[-0.3cm]{0cm}{8mm} Fit A  & $3874.71^{+0.11}_{-0.05} -i~0.28^{+0.04}_{-0.10}$   &  $0.21^{+0.04}_{-0.04}$ & $9^{+11}_{-9}$     &   $0.69^{+0.07}_{-0.12}$ & $9^{+11}_{-8}$    \\
\hline
\rule[-0.3cm]{0cm}{8mm} Fit B  & $3874.75^{+0.12}_{-0.06} -i~0.34^{+0.06}_{-0.12}$   &  $0.23^{+0.03}_{-0.04}$ & $10^{+13}_{-9}$     &   $0.70^{+0.03}_{-0.03}$ & $11^{+13}_{-8}$    \\
\hline
\end{tabular}
\caption{\label{tab:poles;all} The pole locations and their residues (both magnitudes and phases) from our fits, in RS-II.  }
}
\end{center}
\end{table}
As shown in Fig.~\ref{Fig:poletraj;all}, the pole trajectories of Fits A and B are very similar to that of the nominal fit.
\end{document}